\documentclass[letterpaper, 10 pt, conference]{IEEEtran}  

\IEEEoverridecommandlockouts                              
\usepackage{graphics} 
\usepackage{epsfig} 
\usepackage{mathptmx} 
\usepackage{times} 
\usepackage{amsmath} 
\usepackage{amssymb}  
\usepackage{enumerate}
\usepackage{balance}
\usepackage{flushend}
\usepackage{tikz}
\usepackage{cite}
\usepackage{graphicx}
\usepackage{multirow}
\usepackage{hhline}
\usepackage{booktabs}
\usepackage{tabularx,ragged2e}
\usepackage{graphics}
\usepackage{tabularx}
\usepackage{hyperref}
\usepackage{subcaption}
\usepackage{amsfonts}
\usepackage[colorinlistoftodos]{todonotes}
\usepackage{authblk}

\hypersetup{
    colorlinks=true,
    linkcolor=blue,
    filecolor=magenta,      
    urlcolor=blue,
}

\title{\LARGE \bf The Sky is NOT the Limit Anymore: \\
Future Architecture of the Interplanetary Internet
}

\author[*]{Ahmad ALHILAL$^{*}$, Tristan BRAUD} 
\author[$^{*\dag}$]{Pan HUI}

\affil[*]{The Hong Kong University of Science and Technology - Hong Kong}
\affil[$\dag$]{University of Helsinki - Finland}
\affil[]{Email: aalhilal@ust.hk, braudt@ust.hk, panhui@cse.ust.hk
}

\begin{document}

\maketitle

\begin{abstract}
Fifty years after the Apollo program, space exploration has recently been regaining popularity thanks to missions with high media coverage. 
Future space exploration and space station missions will require specific networks to interconnect Earth with other objects and planets in the solar system. 
The interconnections of these networks form the core of an Interplanetary Internet (IPN). 
More specifically, we consider the IPN as the combination of physical infrastructure, network architecture, and technologies to provide communication and navigation services for missions and further applications.
Compared to the current implementation of the Internet, nodes composing the core of the IPN are highly heterogeneous (base stations on planets, satellites etc.). Moreover, nodes are in constant motion over intersecting elliptical planes, which results in highly variable delays and even temporary unavailability of parts of the network.
As such, an IPN has to overcome the challenges of conventional opportunistic networks, with much higher latency and jitter (from a couple of minutes to several days) and the additional constraint of long-term autonomous operations.
In this paper, we highlight the challenges of IPN, demonstrate the elements to deploy within the areas of interest, and propose the technologies to handle deep space communication. We provide recommendations for
an evolutionary IPN implementation, coherent with specific milestones of space exploration.
\end{abstract}

\section{Introduction}


Exploring the world beyond its physical boundaries is one of humanity's oldest dreams.
In recent years, space exploration has not only been feeding human curiosity, but also allowed for scientific advancement in environmental research, and to find natural resources~\cite{daniel2015whyexp,mickael2006whyexp,nasa_iseng2013expbenefit}. Although media exposure reached its peak during the Apollo programs, space research remains a very active domain, with new exploration and observation missions launched every year.

Following the Apollo program, NASA launched Voyager 1 and 2 in 1977\cite{nasa1977voyager1,nasa1977voyager2,nasa1977voyager} to explore Jupiter, Saturn, Uranus and Neptune.
In September 2007, Voyager 1 crossed the termination shock (where the speed of the solar wind drops below the speed of sound) at 84\,AU which is more than twice the distance to Pluto. 
The Voyager Interstellar Mission (VIM), an extension to the 1977 Voyager mission, will explore the outermost edge of the Sun's domain and beyond.
Regarding nearby planets, NASA's twin robot geologists, the Mars Exploration Rovers (MER), landed on Mars in 2004 to perform on-site geological investigations, joined by NASA's Mars Curiosity Rover in 2012\cite{nasa2012marsrovers}. 
Currently, 6 active satellites orbit around Mars, with primary purpose of studying the atmosphere, relay data for other missions such as the Mars rovers, or test key technologies for interplanetary exploration ~\cite{marsodyssey,marsexpress,maven,mom,mro,tgo}.
Deep Impact mission is another mission by NASA which began in 2005 on comet Tempel 1 to expose materials on its surface\cite{nasa2005deepimpact}.
Private companies are also starting to take part in space exploration. In 2012, Red Bull sent the Stratos capsule into the stratosphere, which allowed detailed study of the effect of breaking the sound barrier on a human body. SpaceX and Blue Origin were founded with main objective to reduce the cost and increase the safety of space flight. Finally, NASA recently contracted Boeing to transport astronauts to the International Space Station (ISS)~\cite{boeing}. 

\begin{figure}[t]
\centering
\begin{subfigure}[b]{0.175\textwidth}
\includegraphics[width=0.8\textwidth]{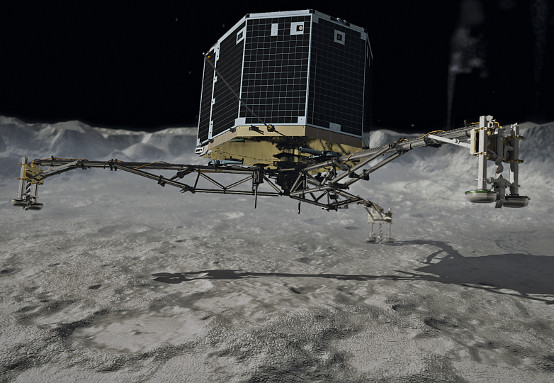}
\subcaption{}
\end{subfigure}
\begin{subfigure}[b]{0.125\textwidth}
\centering
\includegraphics[width=0.8\textwidth]{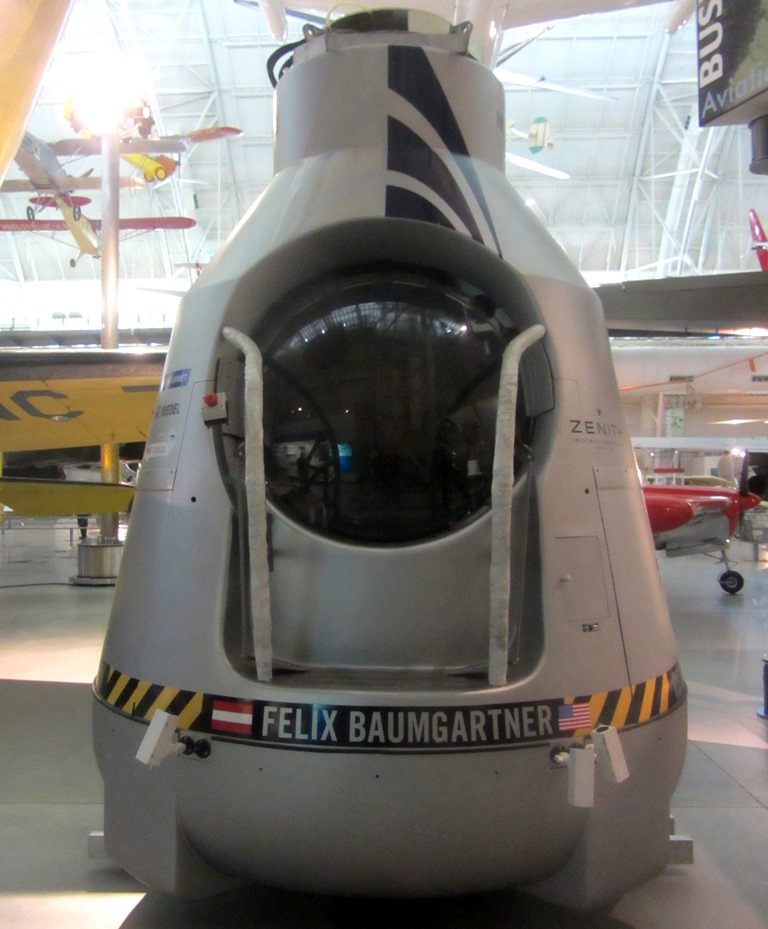}
\subcaption{}
\end{subfigure}
\begin{subfigure}[b]{0.135\textwidth}
\centering
\includegraphics[width=0.7\textwidth]{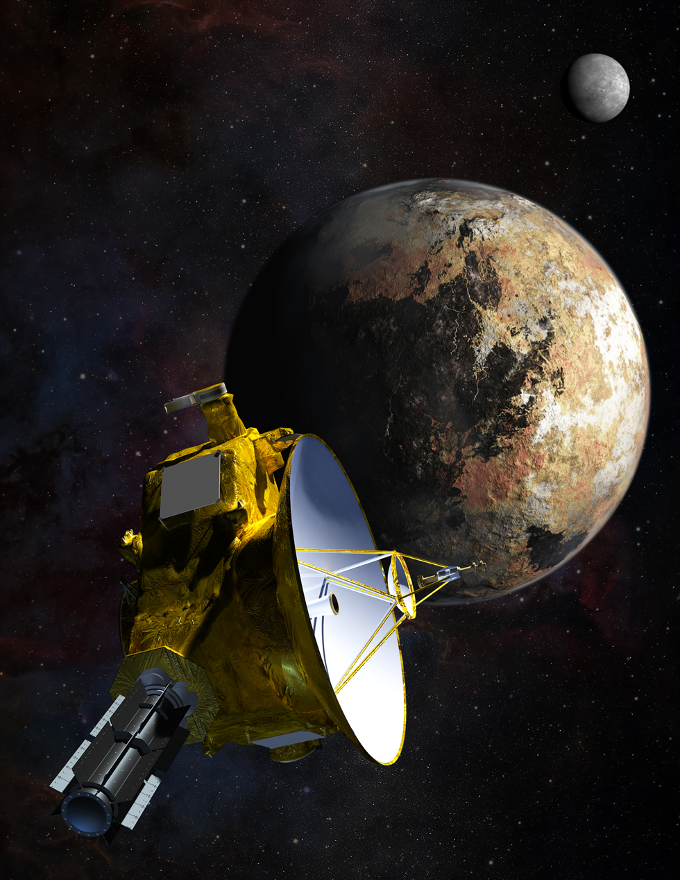}
\subcaption{}
\end{subfigure}
\caption{Recent missions with high media coverage. (a) Philae landing on Rosetta\protect\footnotemark. (b) The Stratos capsule\protect\footnotemark. (c) New Horizon reaching Pluto\protect\footnotemark.}
\label{fig:intro}
\end{figure}
\footnotetext[1]{NASA/Public Domain}
\footnotetext[2]{User:FlugKerl2/Wikimedia Commons/CC BY-SA 4.0}
\footnotetext[3]{DLR/CC BY-SA 3.0}

In recent years, both space agencies and private companies managed to attract a lot of media attention on ongoing and future space missions, resulting in a clear regain of interest of the public for space exploration. 
Figure~\ref{fig:intro} presents several missions with high media coverage. From left to right: Philae landing on the comet Rosetta, the Red Bull Stratos capsule, and New Horizon approaching Pluto. 
As a consequence of this regain of interest, space agencies started to plan several ambitious missions over the $22^{nd}$ century.
The geographic proximity of Mars to Earth, associated with advances of technology enable to envision sending manned missions in the next decades. 
NASA~\cite{journey} is planning round trips to Mars by 2030, while the Mars One project aims at setting a colony by 2032~\cite{marsone}. 
Finally, in a recent declaration, Elon Musk announced that SpaceX's goal is to send the first humans to mars by 2024~\cite{natgeo}.
Several unmanned missions are also planned in upcoming years: new Mars rovers, the James Webb Space Telescope, and even the Asteroid Redirect Robotic Mission(ARRM), a robotic mission to visit a large near-Earth asteroid, collect a multi-ton boulder from its surface, and redirect it into stable orbit around the Moon\cite{nasaarrm}.

Such missions require a reliable, scalable and easy to deploy common communication infrastructure to transmit scientific data from outer space to the earth and back. The advantages of such strategy are manifold:
\begin{itemize}
\item \textbf{Interoperability}: multiple organizations are currently sending spacecrafts in near or deep space. A common infrastructure would significantly reduce the cost of communication, while facilitating inter-agency cooperations. Both manned and unmanned missions would benefit from sharing resources, as it is done currently on the Internet, rather than multiplying parallel incompatible networks.
\item \textbf{Security}: The proliferation of communication systems considerably increases the risk of failure. Due to their critical nature, space communication systems present a high need for redundancy and security. A common network architecture would enable organizations to work together towards a reliable and secure infrastructure.
\item \textbf{Increased bandwidth}: New Horizon's high-resolution pictures of Pluto were a huge success~\cite{newhorizonpluto}, participating in the general public's regain of interest in space exploration. To keep this interest going, future missions will require higher bandwidth to continue providing data to the general audience.
\item \textbf{Scalability}: Space exploration is an incremental process. If we look at Mars exploration, it started with several probes taking pictures, followed by the launch of satellites, and finally robots. Each of these steps requires increasing communication resources. Therefore, it makes more sense to progressively scale up the network than deploying all the resources for future missions at once.
\item \textbf{Colonies}: A round trip to Mars would take 18 months, requiring the astronauts to remain 3 months on Mars, waiting for optimal conditions for the return trip~\cite{marsduration}. Several organizations even envision definitive Mars colonies. Interconnecting both planets networks would facilitate the expansion of human knowledge and culture.
\end{itemize}

In this article, we propose a future Interplanetary Internet (IPN)~\cite{burleigh2003interplanetary} architecture. Such an architecture interconnects networks from various organizations to form a unified network, itself connected to the current Earth Internet. In the long run, an IPN architecture also facilitates the network expansion.
With the IPN Internet being the future Internet that interconnects the solar system, and potentially beyond, we focus on the aspects and challenges involved in such infrastructure.
We propose an effective infrastructure aligned with existing technologies to cope with the foreseen challenges, and reliable protocols to provide autonomous data delivery from an area of interest (planet, Moon) to the Earth and back.

Throughout this paper, we follow a bottom-up approach, from the transmission channel to our proposed architecture. After a quick recap of related work towards an IPN in Section~\ref{sec:related}, we introduce the characteristics of the medium and study the opportunities for transmission in Section~\ref{sec:medium}. We then review the existing infrastructures that can be used in our architecture in Section~\ref{sec:satellite}. We finally propose an evolutionary architecture for Interplanetary Internet in Section~\ref{sec:arch}.
\section{Related Work}
\label{sec:related}

The IPN describes the set of
communication services for scientific data delivery and the navigation services for exploration spacecrafts and orbiters~\cite{bhasin2001advanced,fang2004performance}.
As the IPN is still in its incubation stage, its architecture and infrastructure should be carefully planned; a considerable amount of common standards and research is required to reach an agreement between organizations and cope with the high deployment costs~\cite{mukherjee2015interplanetary}.

The Jet Proportion Laboratory (JPL) deployed the first DTN Gateway located about 25 million kilometers from the Earth during the "Deep Impact Network Experiment (DINET)" \cite{wyatt2009disruption,cerf2009first}. During this experiment, about 300 images were transmitted from the JPL nodes to the spacecraft. These images were then automatically forwarded back to the JPL.
NASA and JPL have started implementing technologies such as the deep space network (DSN), which consists of an array of antennas to cover the whole solar system and terminals for optical communication~\cite{leslie2015realizing,stephen2015realizing}. This network was notably used to retransmit the images of the first moonwalk.

NASA has also conducted studies to provide a common infrastructure for forthcoming space missions. Bhasin et al.~\cite{bhasin2004developing} propose a scalable communication architecture to maximize data delivery and provide capabilities to send high volumes of data ($>100Mb$). This architecture provides high bandwidth communication for further scientific missions.
In this study, the authors define the requirements to achieve such architecture as follows: 
\begin{itemize}
\item The \textbf{architecture elements and interfaces} define the links to connect local planets with a remote in deep space. 
\item A \textbf{layered/integrated communication architecture} should provide end-to-end autonomous data routing. 
\item \textbf{Communication nodes} (rovers, satellites, spacecrafts etc.) take care of the transmission.
\end{itemize}

Another work by NASA and JPL focuses on autonomy to reduce the dependency on resource scheduling provided by Earth operators and increase fault tolerance~\cite{bhasin2001advanced}. This study aims at providing a communication architecture specifically for Mars exploration, defined as a "near-term architecture". The authors also further develop concepts raised in~\cite{bhasin2004developing} such as the Radio Frequencies (X,Ka-band) for Earth to Mars communication, the satellites (MarSat, AMO, etc.) required to maximize bandwidth and get high data rates, and other technologies for Backbone network and proximity networks.

In this paper,  we develop an evolutionary IPN architecture that takes into consideration NASA's needs and requirements\cite{bhasin2004developing,bhasin2001space,bhasin2004evolutionary,bhasin2001advanced,rash2005internet}, but also the time frame for future missions.
 \section{Characteristics of the Space Medium}
\label{sec:medium}

Towards an IPN Internet, we first focus on the transmission of data over the different communication mediums.
Due to the diversity of the propagation mediums in an IPN context, a signal can experience many impairments:~\cite{stallings2002wireless}:
\begin{itemize}
\item \textbf{Attenuation} is a function of frequency. For high frequencies, the signal may experience distortion for large distances. It is, therefore, crucial to amplify high frequency signals.
\item \textbf{Free Space Loss (FSL)} is the primary factor for signal loss which is calculated by the following formula~\cite{saleh1946rfpropagation,friis1946note,stallings2002wireless,rappaport1996wireless}:
\begin{equation}
L_s=(\frac{\lambda}{4\pi R})^2,~\lambda=c/f
\label{eq:fsl}
\end{equation}
Where $R$ is the distance of the link, $\lambda$ the wavelength, $c$ the speed of light and $f$ the signal frequency. 
\item \textbf{Noise} can be thermal, inter-modulation, crosstalk or impulse noise and mixes with the transmitted signal.   
\item \textbf{Delay.} There are four major types of delays \cite{bartolacci2012research,cn2017james}:
Processing delay, storage delay, transmission delay, and propagation delay. In the context of space communication, the propagation delay caused by the long distances is the main challenge.
\item \textbf{Atmospheric Absorption.} On Earth, the peak attenuation occurs in presence of water vapor around 22 GHz, and in presence of oxygen around 60GHz. Other planets have a different atmosphere composition and therefore different absorption to take into account~\cite{mahaffy2013abundance}.
\end{itemize}

\begin{figure}[t]
    \centering
    \begin{subfigure}{0.5\textwidth}
    \centering
    \includegraphics[width=0.9\textwidth]{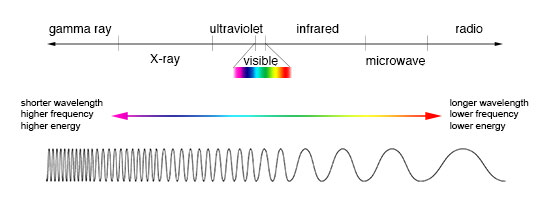}
    \caption{The Electromagnetic spectrum \cite{seeds2017horizons}.}
    \label{fig:elec_spectrum}
    \end{subfigure}
	\begin{subfigure}{0.5\textwidth}
    \centering
    \includegraphics[width=0.9\textwidth]{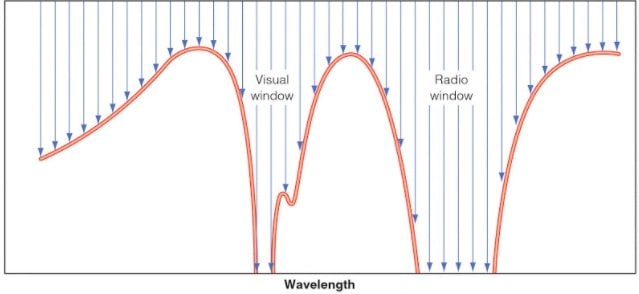}
    \caption{Blocked and passed radiations in the Earth atmosphere (atmospheric windows) 
    \cite{seeds2017horizons}. Only visible light and radio frequencies can propagate in the atmosphere without distortion or absorption.}
    \label{fig:blockedradiation}
   	\end{subfigure}
\end{figure}

Figure~\ref{fig:elec_spectrum} represents the different electromagnetic radiations: radio (RF), infrared (IR), visible, ultraviolet, x-ray, and gamma ray. 
Most of the electromagnetic radiations emitted by outer space don't reach the surface of the Earth except at a very few wavelengths, such as the visible spectrum, also called \textbf{free space optic (FSO)}, \textbf{radio frequencies (RF)}, and some ultraviolet wavelengths, as shown in Figure~\ref{fig:blockedradiation}.  These bands are called atmospheric windows\cite{seeds2017horizons,lujan1994human,nasa2013emspectrum}. Although Earth's atmosphere blocks other bands such as gamma rays, infrared or X-rays, we may utilize them to transmit information in space and through the atmosphere of other planets.
The lower power consumption, lower mass, higher range and higher bandwidth of optical communication (FSO) compared to RF make optical communication the auspicious technology to serve as a communication medium in IPN \cite{kaushal2017optical,williams2007rf,hemmati1997comparative}. 

In practice, NASA's Laser Communication Relay Demonstration (LCRD) mission\cite{lcrd2017} continues the legacy of the Lunar Laser Communications Demonstration (LLCD)\cite{nasa2013llcd}, using FSO.
This latter mission flew aboard a moon-orbiting spacecraft called LADEE, Lunar Atmosphere and Dust Environment Explorer\cite{ladee2013}, in 2013. 
Overall, compared to traditional communications systems on spacecraft, LLCD used half the mass, 25 percent less power, and transmitted six times as much data per second. 
The LCRD project is currently under validation, and its launch is scheduled within a commercial satellite for 2019~\cite{lcrd}.

\section{Space Communication Infrastructure}
\label{sec:satellite}

In an IPN context, satellites constellations serve as an access network to the planetary surface network. In our architecture, the main purpose of satellites is relaying and amplifying radio signals around the curve of the earth. In this section, we present the technologies currently deployed that can serve for the early deployments of an IPN Internet.
    

\subsection{Satellite Orbits and constellations}

\begin{figure}[t]
    \centering
    \includegraphics[width=9cm,height=9cm,keepaspectratio]{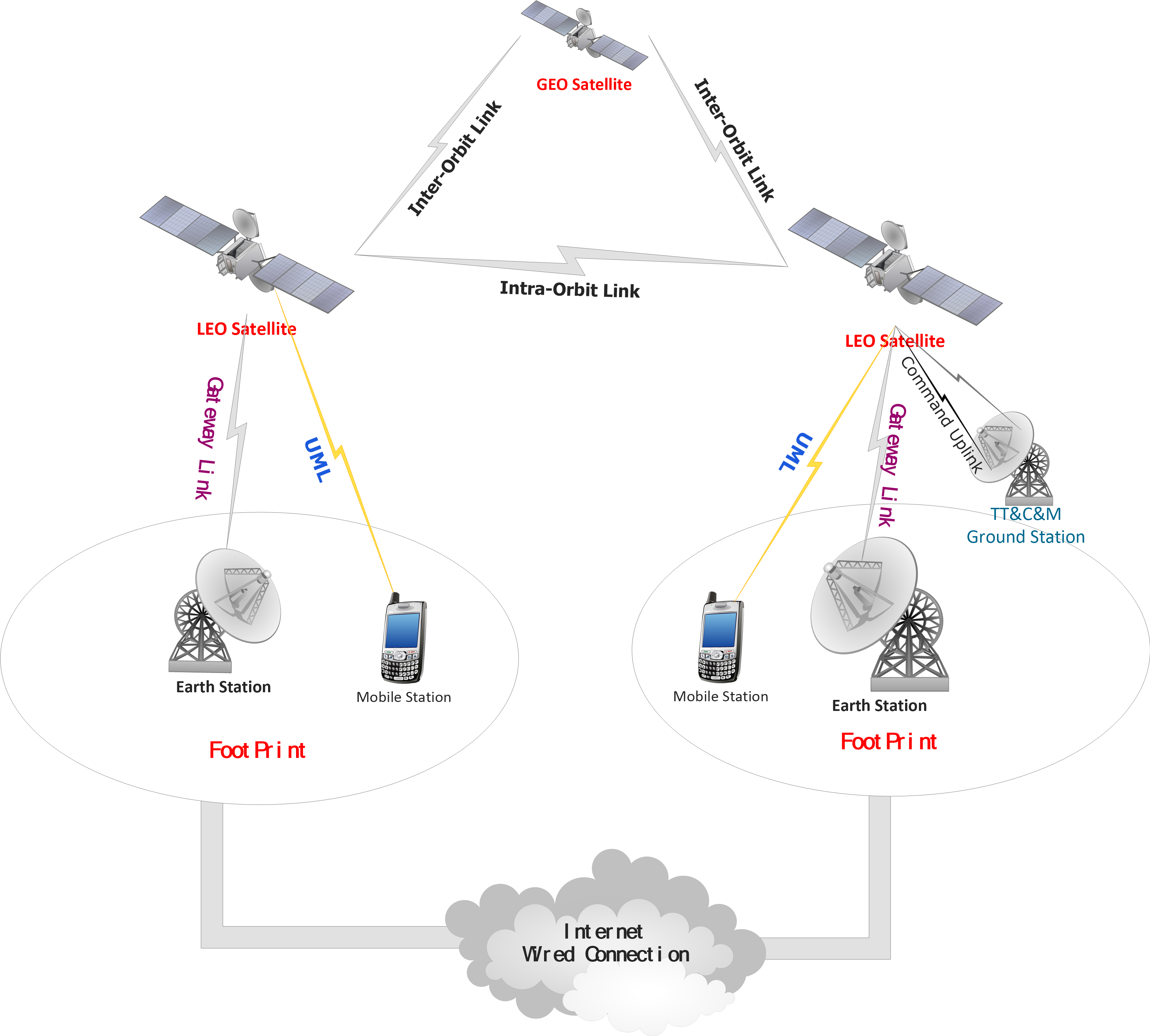}
    \caption{Satellite communication \& architecture. The space segment and the earth segment are interconnected to provide a complete earth coverage.}
    \label{fig:satellitesarch}
\end{figure}

The current satellite infrastructure is presented Figure~\ref{fig:satellitesarch}. This infrastructure can be broken down into two segments~\cite{ippolito2017satellite}: the space segment and the ground (or Earth) segment.
The \textbf{Space Segment} includes the orbiting satellites, and the ground stations that provide the operational control of the satellite(s) in orbit. The \textbf{Ground Segment} consists of the earth surface terminals that employs the communications capabilities of the Space Segment.
A full constellation of satellites is required to cover the surface of the earth. Nowadays, satellites are distributed over three orbits~\cite{stone2004introduction,elbert2008introduction}: 
\begin{itemize}
\item \textbf{Geostationary Earth Orbit (GEO)} are synchronized with the earth rotation and have a 24-hour view of a particular area. 
\item \textbf{Medium Earth Orbit (MEO)} orbit the Earth between 8,000 and 18,000km. As such, they are only visible for a period of 2-8 hours for a specific area. 
\item \textbf{Low Earth Orbit(LEO)} orbit closer to the earth and are only visible for a short period of time (15-20 minutes each pass) in a given area. 
\end{itemize}

The combination of these satellites covers the whole surface of Earth thanks to Inter-Satellite Links (ISL). These links can either connect two satellites on the same orbits or on different orbits~\cite{miller1993satellite}. 
This infrastructure is very convenient for an IPN Internet and can be used as an access network between the Earth and other planets. In the case of colonies on other planets, such constellations could also be deployed to provide full surface coverage and interconnect both planets Internets.

\subsection{Space Communication and Navigations Networks}


The current space communication architecture operated by NASA embraces three operational networks that collectively provide communications services to supported missions using both space and ground-based assets\cite{mukherjee2013communication}:
\begin{figure}[t]
     \centering
     \includegraphics[width=0.4\textwidth,keepaspectratio]{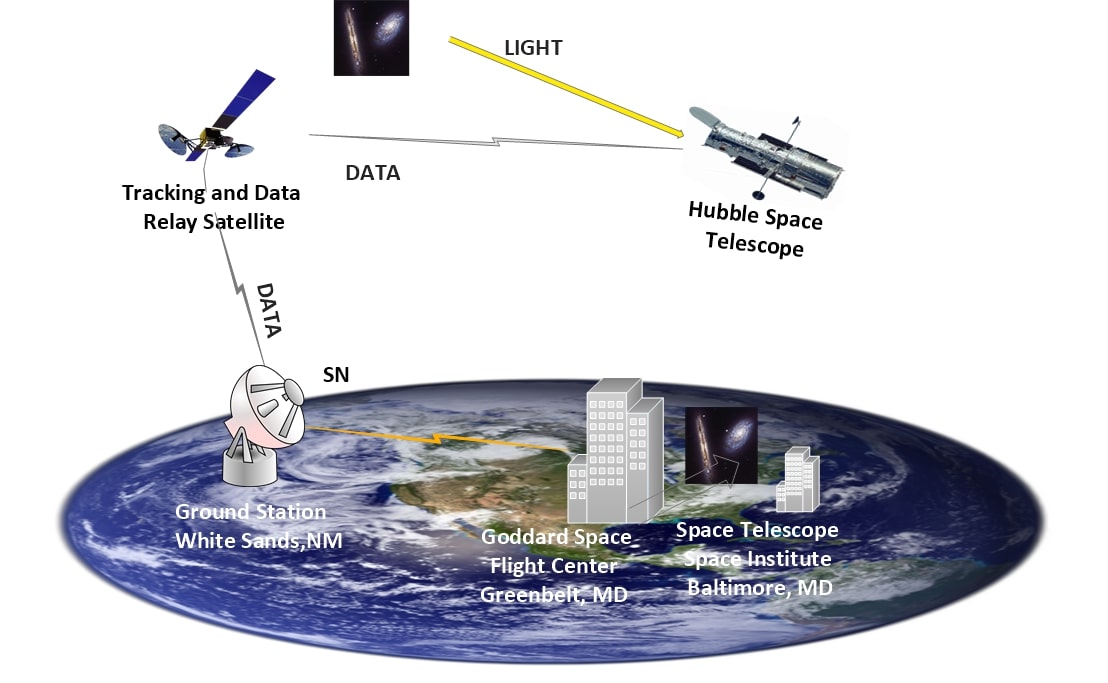}
     \caption{Hubble Observations' Data Path}
     \label{fig:hubbledatapath}
\end{figure}
The \textbf{Deep Space Network (DSN)} is composed of three equidistant ground stations to provide continuous coverage of GEO orbits, and unmanned spacecraft orbiting other planets of our solar system~\cite{nasa2013dsn,renzetti1971dsn}.
The \textbf{Near Earth Network (NEN)} consists of
both NASA and commercial ground stations. 
It also integrates systems providing space communications and tracking services to orbital and suborbital missions.
Finally, the \textbf{Space Network (SN)} is a constellation of geosynchronous relays, Tracking and Data Relay Satellite System (TDRSS), and associated ground stations.
For instance, the TDRSS transmits the observations of the TDRSS as shown Figure~\ref{fig:hubbledatapath}.
These three networks are currently deployed in parallel with minimal interaction. The first step towards an extended IPN would be to chain relays located near-Earth, in the solar system, and in deep space altogether. 

\subsection{Towards Interplanetary Communication}

Most of the nodes involved in an IPN are revolving around other stellar objects: planets revolve around the Sun with long distances, satellites orbit planets at a relatively close range. This motion poses many challenges for interplanetary communication\cite{akyildiz2004state,akyildiz2003interplanetary,durst2000not}:
\begin{itemize}
\item \textbf{Extremely long and variable propagation delays}: 3 to 20 minutes from Mars to Earth, 4 to 7 hours from Pluto to Earth, depending on the relative positions of the planets.
\item \textbf{Intermittent link connectivity}: the Sun or other planets may temporarily obscure a given link between two stellar objects. For instance, the Earth to Mars line of sight is regularly obstructed by the Sun when they reach the opposite position in their orbits.
\item \textbf{Low and asymmetric bandwidth}: the limited payload available on satellites severely impacts their transmission power compared to Earth transmission relays.
\item \textbf{Absence of fixed infrastructure}: the IPN is a purely opportunistic network, as the relative position of two nodes is in constant movement with great variations in the link characteristics except for Geocentric satellites.
\end{itemize}

An IPN must address these constraints to optimize the few resources available in the system.
On the other hand, contrary to many other opportunistic networks, the movement of potential nodes in the solar system follows regular patterns that can be predicted to some extent. Moreover, the number of stellar objects enables us to find a path between two nodes even in the case of obstruction, using other nodes as relays. 

\subsection{Delay Tolerant Networks (DTN)}

The IPN Internet is an opportunistic network in which end-to-end latencies can reach up to a day, and jitter is measured in hours. 
As such, conventional Internet architectures based on the TCP/IP stack are not applicable. 
On the other hand, delay-tolerant architectures and protocols were designed to withstand the extreme constrains of the system. 
The first concepts of Delay Tolerant Networks (DTN)~\cite{fall2003delay,burleigh2003delay,fall2007delay} were originally proposed to cope with the characteristics of deep space communication (long delays, discontinuous network connectivity) before being extended to other domains.
The nodes of a DTN infrastructure are called DTN gateways and provide store-and-forward capabilities to handle the eventual link unavailabilities.
To cope with the long and variable latencies, DTN architectures insert
an overlay network protocol called \textbf{Bundling Protocol (BP)}, that provides end-to-end transmission between heterogeneous links. This overlay takes place on top of the transport protocol, ensuring compatibility with existing Internet infrastructures. At each point of the network, BP employs the transport protocol adapted to the transmission conditions.  BP can therefore operate over TCP~\cite{fall2011tcp,postel1981transmission} and UDP~\cite{postel1980user}, but also the Licklider Transport Protocol (LTP)~\cite{ramadas2008licklider,burleigh2008rfc,farrell2008licklider}, a  point-to-point transmission protocol for intermittent links with long propagation delays. Such characteristics make it particularly suitable for interplanetary transmission, where the traditional TCP/IP paradigm cannot be applied. There are many DTN implementations for the Bundling Protocol (BP) that provides the store-and-forward capabilities required in deep space environment. In particular, Interplanetary Overlay Network (ION), and IRB-DTN were designed for deep space communication. 
\section{Architecture \& Communication Infrastructure  for future IPN}
\label{sec:arch}
The ever-increasing interest to explore the universe motivated us to provide an evolutionary architecture that tackles the IPN challenges and bridge large distances in space. After a short summary of the main challenges, we detail our vision for a long-term reliable and scalable IPN architecture.

\subsection{IPN Challenges}
\noindent Deep space networking presents critical challenges~\cite{akyildiz2004state,akyildiz2003interplanetary,durst2000not}:
\begin{itemize}
\item \textbf{Distance between planets:} extremely long propagation delays and high path attenuation between nodes.
\item \textbf{Planetary motion:} not only are propagation delays high but due to planetary motion, they are extremely variable. Planetary motion also leads to intermittent link connectivity due to conjunction or obstruction\cite{nasa2017conjunction}.
\item \textbf{Low embeddable payload}: satellites can only carry a limited payload, which forces us to focus on power, mass, size, and cost for communication hardware and protocol design. Asymmetric bandwidth on the order of 1000:1 is a  direct consequence of this limited payload.
\item \textbf{Relative inaccessibility}: the distance between planets leads to long travel distances. In order to safely launch a mission to a given planet, space travel can only take place on specific days, corresponding to favorable conjunctions. As such, an IPN architecture should focus on backward compatibility and scalability in order to reduce the time and cost of deployment.
\end{itemize}

With these challenges in mind, we propose an evolutive architecture specifically designed to withstand the harsh constraints of deep space communication.

\subsection{IPN Infrastructure}

Due to the high complexity of deployment, we define 3 types of intercompatible architectures, each corresponding to different milestones in space exploration. We refer to them as near-term (current missions), mid-term (human colony on Mars) and long-term architectures (manned and unmanned colonization of the complete solar system).

\subsubsection{IPN Near-Term Communication Architecture}~\\
\indent We propose an IPN near-term communication architecture for the current missions targeting Mars and the Moon. Indeed, both are accessible within a reasonable amount of time (3 days to the moon and 6 to 9 months to  Mars) and several organizations are already planning manned missions within the next 10 years. This architecture reuses a maximum amount of available technologies to interconnect the Earth, Mars and the Moon in a short time frame.

\begin{figure}[t]
    \centering
    \includegraphics[width=0.45\textwidth]{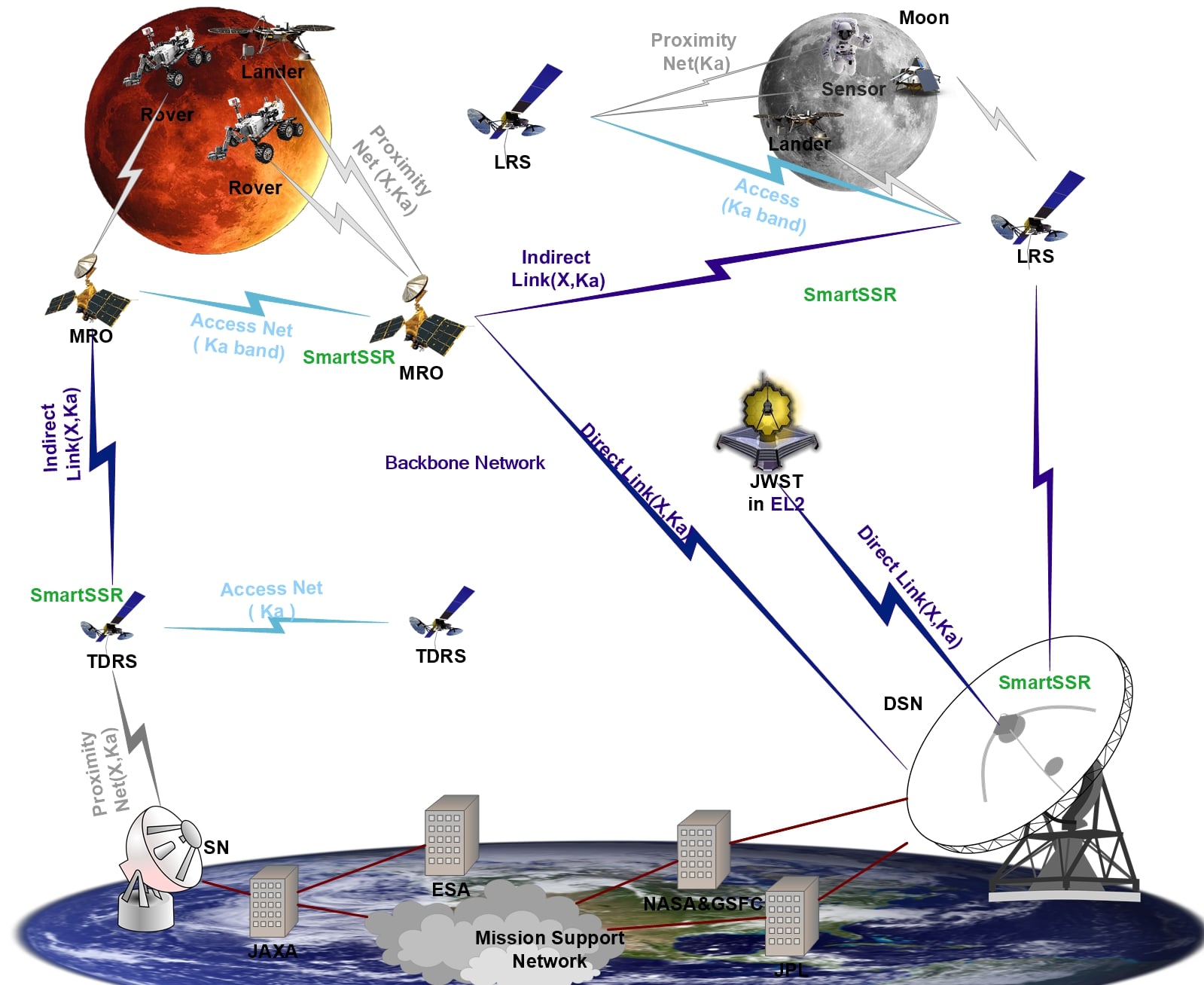}
    \caption{IPN near-term Communication Infrastructure. This architecture reuses the existing infrastructure with minimal addition.}
    \label{fig:ipnnearterm}
\end{figure}

Figure~\ref{fig:ipnnearterm} illustrates the IPN near-term architecture.
We separate this architecture in two sub-systems: the physical layer, that we will call \textbf{Spectrum}, and the upper layers, referred to as \textbf{Network}.  
The \textbf{spectrum sub-system} provides two bands in the microwave carry for data: Ka (26.5-40GHz) and X (8-12.4GHz).
These bands provide higher data rates than the conventional RF bands. The Ka-band allows for the communication in the backbone network and for inter-satellite communication due to its higher frequency (thus, higher data rates). X and Ka together allow for the communication from satellites to the surface of the planet. In our architecture, we switch between both bands depending on the weather as the Ka-band suffers from attenuation in presence of humidity.

The \textbf{network sub-system} contains three sub-networks (see Figure~\ref{fig:ipnnearterm}): The \textbf{proximity network} contains the inter-element links relatively close to the planet or the Moon and the surface networks. The \textbf{access network} consists of satellites orbiting the planet or the Moon interconnected with each other. In this architecture, there are three access networks formed by the satellites orbiting each planet and the Moon. The \textbf{backbone network} interconnects the three access networks with the DSN stations on Earth. 
This network provides two kinds of links for interconnection: direct links and indirect links. 
Direct link connects Mars and lunar relay satellites directly to the DSN on Earth. 
Indirect links go from Mars and lunar relay satellites to Earth relay satellites where the data is then directed to DSN antennas. 
We propose to launch four Lunar Relay Satellites (LRS) and Mars Relay Orbiter (MRO). Three relay satellites are operating, while the fourth remains as a spare. 
We also reuse existing infrastructure. For instance, the Mars Reconnaissance Orbiter satellites are planned to function until 2030 and could be employed in near term.
On Earth, the satellites that serve as data relay are the Tracking and Data Relay Satellites (TDRS). Ten of them are currently operating in geostationary orbit.
From the equipment on a foreign planet to the Earth, each node along the path contains DTN technology to support store-and-forward mechanism. This technology is installed on DSN, relay satellites (LRS, MRO, TDRS), and even in the proximity networks: Landers, Robots, Rovers (data collectors), SN and mission centers (data destination). 

The James Webb Space Telescope (JWST)\cite{gardner2006james,nasa2017jwst} is a large, cold and infrared-optimized space observatory that will be launched into orbit in 2019. This telescope is a good example of a mission that could utilize such architecture. 

\begin{figure}[t]
    \centering
    \includegraphics[width=0.45\textwidth]{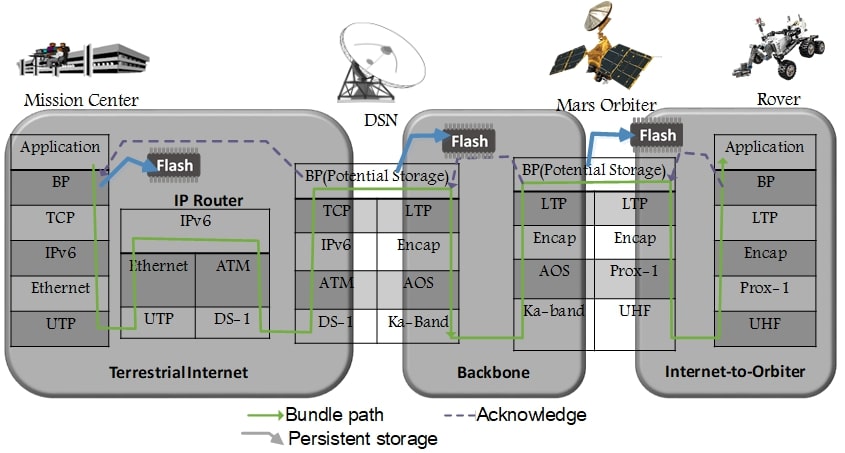}
    \caption{End-to-End Data Transfer Using DTN. In the near-term architecture, the DSN directly connects to the Mars Orbiter, which relays the bundle to the rover using BP over LTP.}
    \label{fig:bundlepath}
\end{figure}

Fig~\ref{fig:bundlepath} outlines the data transmission process from the mission center on the Earth to a Mars rover. The BP is deployed as an overlay on top of TCP between the mission center and the DSN antennas, as TCP is the transport protocol in the terrestrial Internet. In the second trunk of the path, BP operates on top of LTP between the DSN antennas, the Mars Orbiter and the Mars Rover. On this trunk, we transmit data using the microwave band Ka and use LTP due to the long distance -- high latency -- between the Earth and Mars. The green continuous line depicts the data path from the mission center to the Rover. The purple dashed lines show the hop to hop acknowledgments between two neighboring elements.

Currently, the Mars and Moon Orbiters use the Proximity-1 data link protocol\cite{book2013proximity} to communicate with the surface elements and the Advanced Orbiting Systems (AOS) space data link protocol\cite{ccsds2015aosspace} for communication between orbiters and the DSN antennas on Earth.

\subsubsection{IPN Mid-Term Communication Architecture}~\\
\indent For our mid-term architecture, we consider the human colonization of Mars and the further side of the moon, which will extend to colonizing the whole solar system long term. As such, we expect an ever-increasing demand to exchange huge amounts of data in both directions. This phenomenon will be amplified in case of long-term colonies on foreign planets. In this scenario, interplanetary links not only provide Internet access outside of the Earth, but also interconnect the planet Internets in the same way transatlantic cables interconnect continents on Earth.
Future architectures should, therefore, be scalable and tackle the usual bandwidth asymmetry issues to address these challenges.
In our mid-term architecture, we propose using an onboard optical module for spacecrafts and optical communication terminals (OCT) on the planet's surface to support two-way communication with high data rates. This design allows us to considerably reduce the bandwidth asymmetry.
These technologies require less power and considerably reduce the payload. They are also able to reach longer distances and provide higher data rate, 10X - 100X higher than RF.     

\begin{figure}[t]
    \centering
    \includegraphics[width=0.45\textwidth]{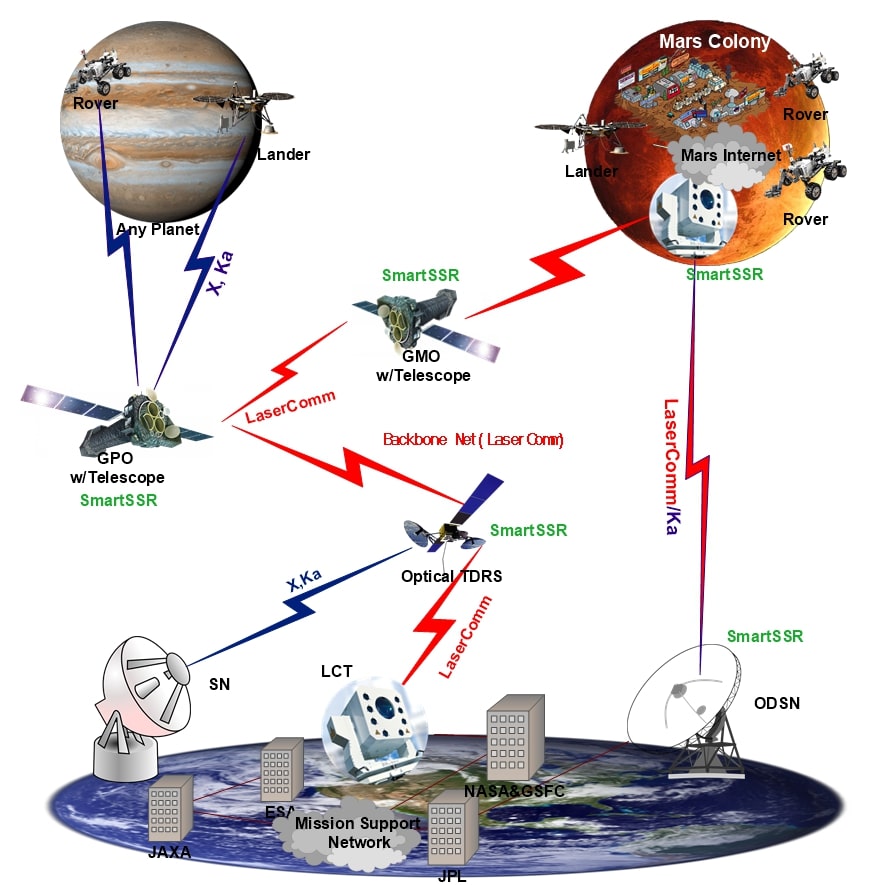}
    \caption{IPN mid-term Communication Infrastructure. We start to deploy Lasercom for long distance links, and extend the architecture to other planets.}
    \label{fig:ipnmidterm}
\end{figure}

Figure~\ref{fig:ipnmidterm} illustrates the IPN mid-term architecture that interconnects the Earth with Mars and other planets. In this architecture, we upgrade the transmission spectrum from microwave (X, Ka) to laser Communication, also referred to as Free Space Optic (FSO). Optical communication is an emerging technology in which data is modulated onto a laser for transmission. The laser beam is significantly narrower than an RF beam and thus promises to deliver more power and achieve higher data rates. 
In outer space, the communication range of FSO communication is in the order of thousands of kilometers. Optical telescopes therefore play a pivotal role as beam
expanders to bridge interplanetary distances of millions of kilometers. Our near-term architecture includes specific hardware to ensure the provision of optical communication. 
It includes constellations of geostationary orbiters or satellites (Optical TDRS around Earth, Geostationary Mars Orbiters(GMOs), Geostationary Planet Orbiter (GPOs)) to provide relay service between nodes on the surface of the outer planet, in-between planets and between the access network from other planets. 
On each satellite, we embed a small (a dozen cm) Cassegrain reflector to support optical communication.
The Optical Deep Space Network (ODSN) substitutes the DSN ground stations by supporting two communication technologies: RF microwaves (X and Ka-band) and optical (Lasercom). This hybrid result in installing optical mirrors in the inner 8m of a standard DSN 34m beam waveguide antenna. The RF communication is kept in order to maintain the operation in all weather conditions. 
The architecture also operates new Laser Communication Terminals (LCT) to exchange data with the GMO, GPO and OTDRS satellites. The LCT contains six small (a dozen cm) refractive telescopes for the transmitter and a single bigger reflective telescope as a receiver. This latter telescope is connected via optical fibers to the respective transmitters and receivers.

\subsubsection{IPN Long-Term Communication Architecture}~\\
\begin{figure}[t]
    \centering
    \includegraphics[width=0.4\textwidth]{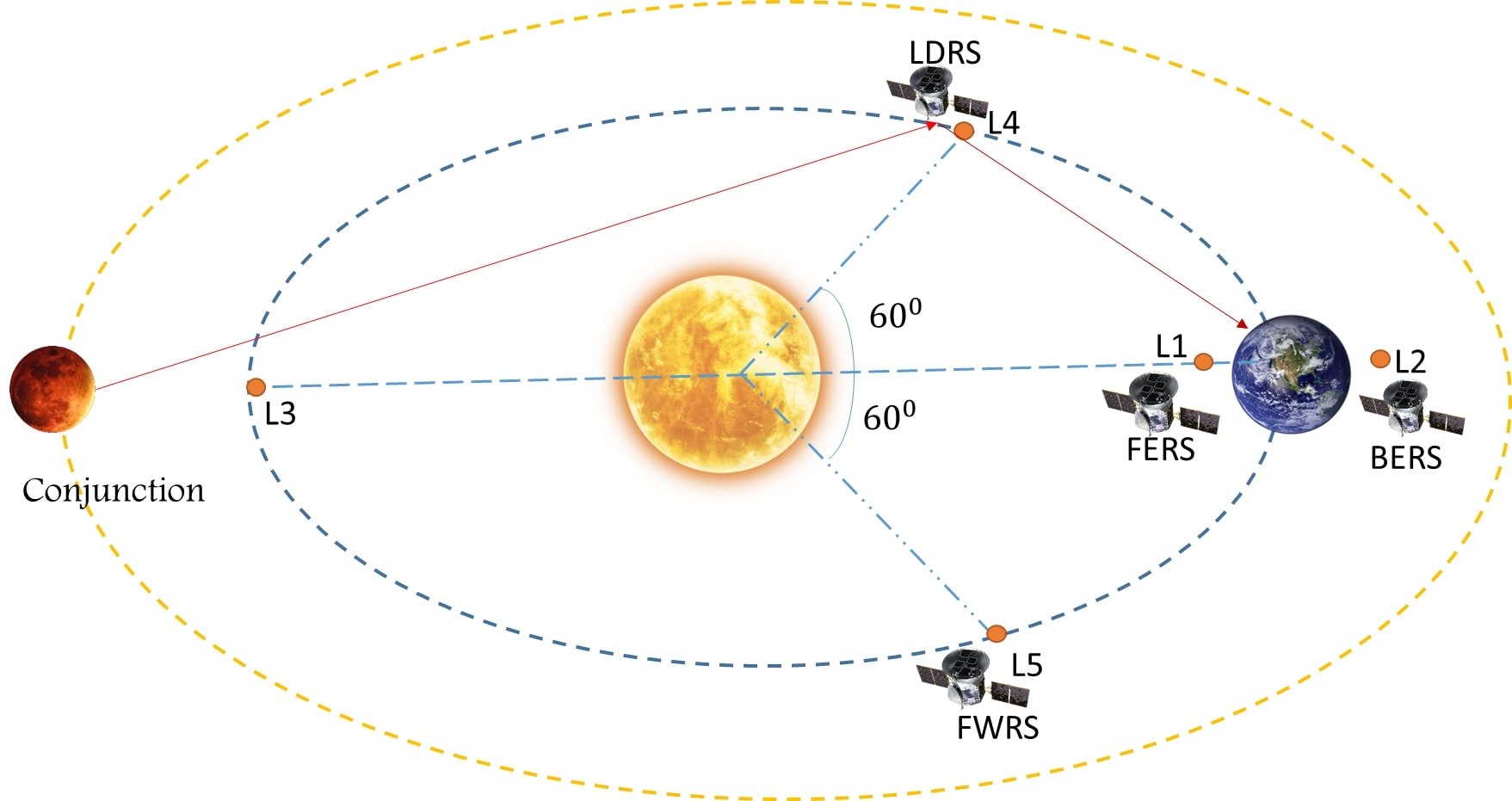}
    \caption{Lagrangian points and spacecrafts placement for Sun-Earth Lagrangian points. When the Mars-Earth LOS is obscured, data can go through the LDRS or FWRS to avoid service interruption.}
    \label{fig:lagrangianpoint}
\end{figure}
\begin{figure}[t]
    \centering
    \includegraphics[width=0.45\textwidth]{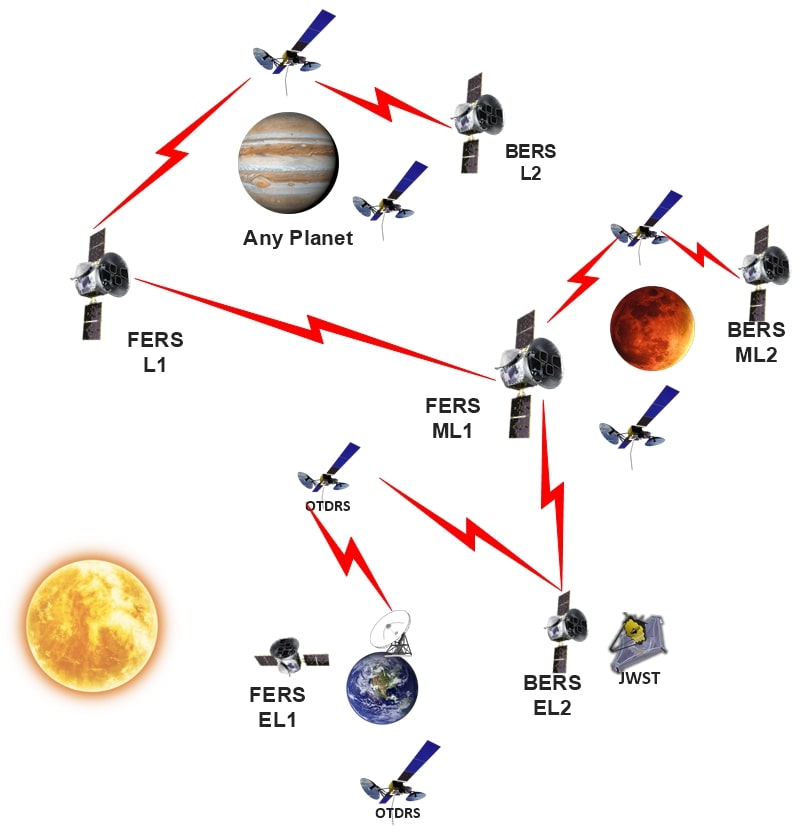}
    \caption{IPN long-term Communication Infrastructure. Lasercom is now generalized to all communication. We increase redundancy with the introduction of more spacecrafts in Lagrangian points as data relays.}
    \label{fig:longterm}
\end{figure}
\indent Optical communication in space is based on line of sight (LOS), which may experience obstruction or conjunction. 
For instance, Earth and Mars can be obscured from each other by the Sun . This obstruction lasts for two weeks every 26 months. Moreover, LOS communication in space attenuates because of Free-Space Loss (FSL) which increases with distances, as we showed in Equation~\ref{eq:fsl}. Therefore, 
communication between the Earth and further planets experiences much more attenuation than communication between the Earth and Mars. If we consider transmission between the Earth and Pluto, the signal travels 38.44\,AU = 5,766,million km (0.52\,AU for Earth to Mars) in space and needs 5.4\,hours to reach its destination.

We propose operating spacecrafts in Sun-Earth's Lagrangian points to address these problems.
Fig ~\ref{fig:lagrangianpoint} shows the positions of the five Lagrangian Points L1,L2,L3,L4,L5. At each point, the gravitational forces of two large bodies (Sun-Earth for instance) cancel the centrifugal force. A spacecraft can therefore occupy the point and move around the Sun without the need for external intervention. These points are commonly used for observation missions and are envisioned as relays for space colonization~\cite{spacecolonization}.
In our architecture, we employ these points to operate spacecrafts as repeater and relay nodes in deep space. 
These nodes bridge the distances between the planets, where they relay data using lasercom to the destination. They also allow to multiply the number of orbiting points in the solar system to propose alternative paths in case of obstruction or conjunction.
More specifically, we propose to place spacecrafts in points L4 and L5 to address obstruction and conjunction, and spacecrafts in L1 and L2 to tackle attenuation.

In the case of a planet-to-Earth transmission, data first goes through a Front End Relay Spacecraft (FERS) in the source's planet Lagrangian point L1. On the relay planet, data then passes through a Front End Relay Spacecraft (FERS) in L1 and a Back End Relay Spacecraft (BERS) in L2. Finally, on Earth, we position a FERS in Earth L1 (EL1), a BERS in EL2, a Lead Relay Spacecraft (LDRS) in EL4 and a Follow Relay Spacecraft (FWRS) in EL5, as shown Figure~\ref{fig:lagrangianpoint}. The LDRS and the FWRS maintain the communication with other planets even if they are located behind the Sun. They also help to fragment the communication distance, especially for the communication with planets on the opposite side of the Sun.

\subsection{Autonomous Routing}

Currently, space communication systems are mission-specific and point-to-point. Moreover, they are dependent on operator-specific resources. 
Our approach aims at reducing the dependency on resource scheduling provided by Earth operators and interconnect the planets. To do so, our architecture provides autonomous operation of the spacecrafts, as well as autonomous routing of commands and data in space. 

We adopt DTN to provide autonomous data routing mechanisms similar to Terrestrial Internet routing capabilities. This approach solves half of the autonomous routing issue by providing an overlay over the lower five OSI layers. As such, the transmission is independent of the underlying routing protocols. However, DTN operates on point-to-point communication, whereas IPN needs multi-point communication, which represents the other half of the autonomous routing issues.
We solve these issues by allowing the routing protocol to control the antenna pointing, transmit power and data rates, and provide synchronization capabilities between the sender or the receiver.
This solution requires interactive links between the nodes that can be created and broken on demand at any time in the whole IPN Network. 
These on-demand features require specific hardware for pointing and focusing transmission. 
We propose using 2-axis gimbals (azimuth and elevation) \cite{guelman2004acquisition,talmor2016two},  Coarse Pointing Assembly (CPA) and Fine Pointing Assembly (FPA)\cite{guelman2004acquisition,talmor2016two,skullestad1999pointing,yamashita2011new,barho2003coarse} to orient the antenna and the beam. 
Beside DTN, we combine the two following mechanisms to provide autonomous data routing and mimic the communication used in mobile networks:
\begin{itemize}
\item \textbf{Locating and Calculation Subsystem.}
 Each Optical Communication Terminal (OCT) computes onboard the orbital position of its partner (receiver OCT) and finds its angular velocity. This is the initial telescope pointing phase. 
This approach uses the reference position and the new position as input parameters for the position controller, and the reference velocity and the current velocity to feed in the velocity controller. This ensures inertial LOS orientation towards the partner for stabilization purposes. The OCT also calculates the distance to the partner and adjusts the transmission power accordingly. 
\item \textbf{Pointing Control Subsystem.} 
This technique employs the same laser for transmitting and as a beacon. The beam-width is controlled from broad in the acquisition stage (also referred to as Coarse Pointing), to narrow in the tracking stage (also referred to as Fine Pointing). The acquisition is achieved by the hardware 2-axis gimbals' pointing and a Coarse Pointing Assembly (CPA) which allow contact with broad beacon beam. When acquired, the beam focusing phase (Fine Pointing) progressively narrows the beam while correcting the pointing accuracy up to sufficient level of beam concentration to get maximum received power, thus high data-rates. This stage uses either the Fine Pointing Assembly (FPA) or beam control approach which includes three control components: Fast Steering Mirror (FSM), Point Ahead Mirror, and Laser Beam Defocus Mechanism (LBD). 
\end{itemize}

We combine both mechanisms to quickly find a new partner and reduce the off-line time.

\subsection{Technologies integration}
\label{sec:technologiesintegration}

We now draw the attention to the technologies that make this architecture possible. We specifically focus on the communication and hardware aspects of a durable infrastructure.

\subsubsection{SmartSSR DTN Router}~\\
\begin{figure}[t]
    \centering
    \includegraphics[width=0.35\textwidth]{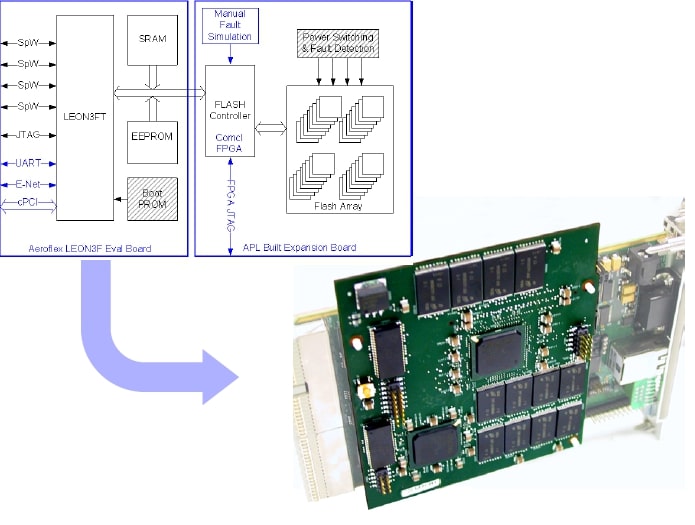}
    \caption{The SmartSSR Prototype Development Board Architecture.}
    \label{fig:smartssr}
\end{figure}
To interconnect the local Internet, with other planets' Internet, a DTN Router  (also referred to as DTN Gateway) is necessary. This router is provided with one of DTN implementation that contains the bundling protocol, convergence layer protocol and transport protocols, LTP, TCP, and UDP. 
DTN Routers should meet the payload constraints of the satellites and spacecraft. 

The SmartSSR is a solid-state recorder (SSR) developed by the Applied Physics Laboratory (APL) that enables spacecrafts to operate as DTN routers. Its small mass and size combined with its relative low cost make it easy to massively install on the payload module of any spacecraft. These features make the SmartSSR an optimal choice to provide DTN capabilities.
The SmartSSR's main goal is to tackle intermittent connectivity. It features JPL's DTN ION implementation to store large amounts of data when the link to the next hop is not available. It also features LTP as a transport protocol for delivering data. The SmartSSR combines a NAND flash array with a general purpose processor to host several closely related functions (see Figure~\ref{fig:smartssr}). This processor is based on LEON3FT processor and uses SpaceWire interfaces to communicate with other spacecrafts components. The SmartSSR uses the Space File System (SpaceFFS) to manage spacecraft data. This FS is adapted of a flash file system to meet the special requirements of space operational environment \cite{krupiarz2011enabling,iaf2010smartssr,mick2011smartssr}.

\subsubsection{Optical Communication}~\\
The key technologies for providing optical communication are the flight terminals, ground laser transmitters and ground laser receivers.
Both the flight terminal and ground receivers use telescopes to operate as beam expanders to bridge interplanetary distances of millions of kilometers. 
The flight laser terminal (FLT) carries a 22cm aperture, 4\,W  laser and contains an isolation and pointing assembly (IPA) for operating in the presence of spacecraft vibrational disturbance, and a photon-counting camera to enable the acquisition, tracking and signal reception.
The ground terminals contain photon-counting ground detectors that can be integrated with large aperture ground collecting apertures (telescopes) for detecting the faint downlink signal from deep space.
In our architecture, we propose to place optical communication terminals (OCT), also referred to as Laser Communication Terminal (LCT), like those used in OCTL.
We also consider embedding Cassegrain telescopes beside FLT as a part of the payload of the orbiter satellites and relay spacecrafts in Lagrangian Points to operate as transceivers and provide the autonomous operations.

\subsection{Integration Scenario}

Let's consider an exploration scenario on Jupiter, with several rovers, sensors, and robots working together. These modules generate a meta-data file containing sensed data directed to Earth.
In this section, we describe the full operation of the system, from the communication medium to the selected path and the involved nodes. In this scenario, we use Mars as a relay between Jupiter and Earth to re-amplify the signal and counteract eventual occlusion.


\begin{figure}
	\centering
	\begin{subfigure}{0.45\textwidth} 
		\includegraphics[width=\textwidth]{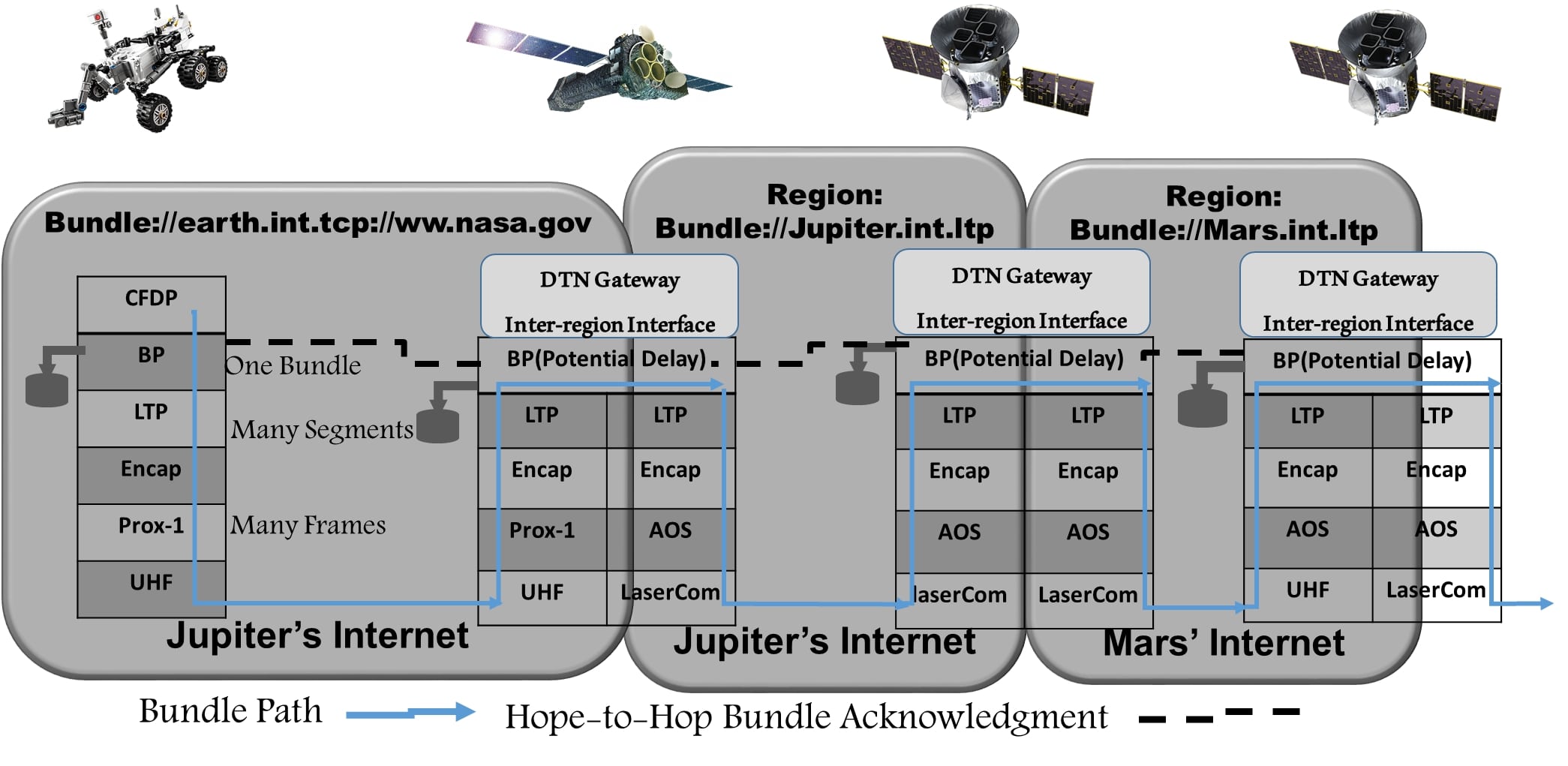}
		\caption{Jupiter Rover to Mars BERS. Data flows through Jupiter's Geostationary orbiter, which transmits to Jupiter's FERS. Jupiter's FERS then transmit the data to Mars BERS.} 
	\end{subfigure}
	\vspace{1em} 
	\begin{subfigure}{0.45\textwidth} 
		\includegraphics[width=\textwidth]{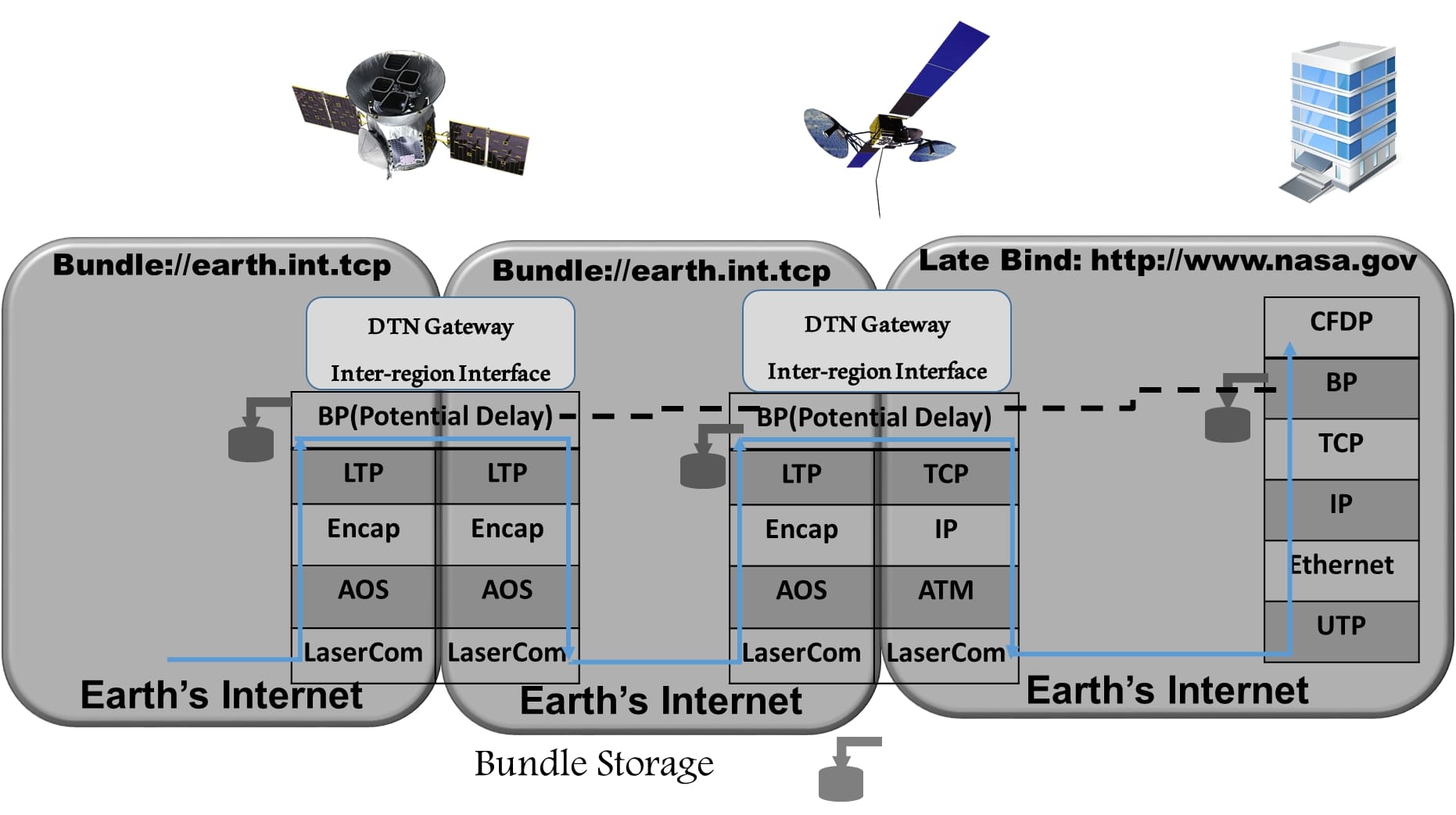}
		\caption{Mars FERS to Earth Mission Center. Mars FERS transmits the bundles either to Mars FERS or to Earth LDRS, depending on conjunction. Earth LDRS then transmits the bundles to the mission center.} 
	\end{subfigure}
	\caption{End-to-End data path for long-term Architecture.} 
    \label{fig:longterm-dtnpath}
\end{figure}

We use SmartSSR in each node to provide DTN capabilities. Each node stores and carries data until a link to the destination is available.
DTN uses tiered naming, addressing and routing. It uses region ID as global unique identifier and entity ID for late binding as follows: \texttt{Bundle://regionID:entitiyID}. A Jupiter rover sending a file to Nasa will name the bundle \texttt{ Bundle://Earth.int.tcp://www.nasa.gov}, with \texttt{Bundle} the transmission unit, \texttt{Earth.int.tcp} the destination regionID, and \texttt{www.nasa.gov} the destination entityID. This naming convention allows the bundle to be first directed to the Earth. Once the bundle reaches the Earth, it is then directed to NASA using the Earth Internet.

We represent the operation of our system in 
Figure~\ref{fig:longterm-dtnpath}. 
The Jupiter rover sends the file using the international communication protocol CCSDS File Delivery Protocol (CFDP). CFDP provides reliable delivery of data and has been specifically designed for use across space links \cite{book2007ccsds,ray2003ccsds}. 
In the region \texttt{Jupiter.int.ltp}, the Bundling Protocol (BP) encapsulates the file into bundles. These bundles are stored on the rover and forwarded when the link with Geostationary Jupiter Orbiter (GJO) becomes available. Once the link to the GJO is available, 
BP invokes LTP to transport each bundle as segments. The segments are then sent to the GJO using either laser communication or Ultra High Frequency RF (UHF).
The GJO then follows the same logic to transmit the bundles to Jupiter's FERS, using laser communication.
Jupiter's FERS will try to communicate with element with it has in its line of sight. If the link to  Mars BERS is not available, Jupiter's FERS stores the bundles on persistent storage (Flash NAND in SmartSSR) and carries the bundles until the link to the next hop becomes available.

Let's assume that Mars is the closest planet between Jupiter and the Earth. Jupiter's FERS will communicate with Mars BERS using the same protocols and medium. The next region is then \texttt{mars.int.ltp}  to which Jupiter's FERS can forward the bundles that it has in its persistent storage. As soon as  Jupiter's FERS receives acknowledgment from Mars BERS, it remove the bundles from the persistent storage. The Mars BERS relays the bundles through Mars FERS, or directly to Earth's LDRS, in the region \texttt{Earth.int.tcp}, based on on-board autonomous routing. 

The Earth's LDRS communicates with the nearest Optical TDRS (OTDRS)
orbiting the Earth to pass the bundles using the same protocols and communication medium.
The OTDRS then communicates with the OCT. 
To send each bundle, BP invokes the underlying convergence layer agent to transform from LTP to TCP to transport the bundles. 
The OCT directs the optics to the control room where the laser is demodulated digital data. 
This data is finally transferred using unshielded twisted pairs cables (UTP) to the mission center, with destination \texttt{www.nasa.gov}. 
After delivering data to mission control center, the de-encapsulation process converts the TCP segments into bundles and deliver them into the CFDP to build the file. 
The BP in the mission center's DTN node creates an acknowledgment to confirm data reception. This acknowledgment follows the backward path to confirm the delivery of data on hop-to-hop basis (dotted lines on Figure~\ref{fig:longterm-dtnpath}).
In this system, the transmission protocols (LTP and TCP) ensure DTN node-to-node reliability. BP provides end-to-end reliability.

\subsection{IPN Implementation Notes}

\begin{figure}
	\centering
	\begin{subfigure}{0.45\textwidth} 
		\includegraphics[width=\textwidth]{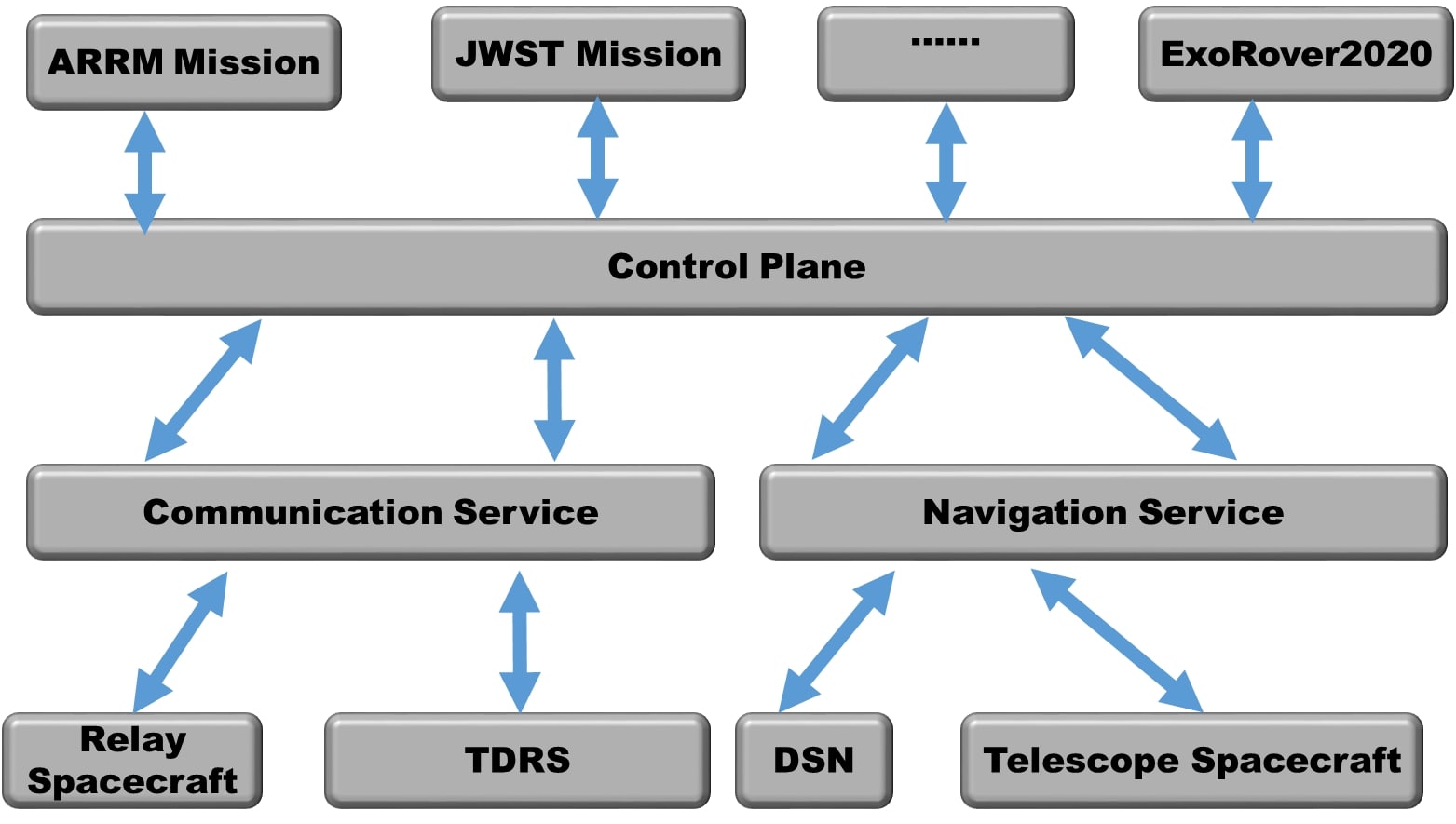}
		\subcaption{Mission-independent SDN hierarchy over the whole IPN.} 
	\end{subfigure}
	\vspace{1em} 
	\begin{subfigure}{0.45\textwidth} 
		\includegraphics[width=\textwidth]{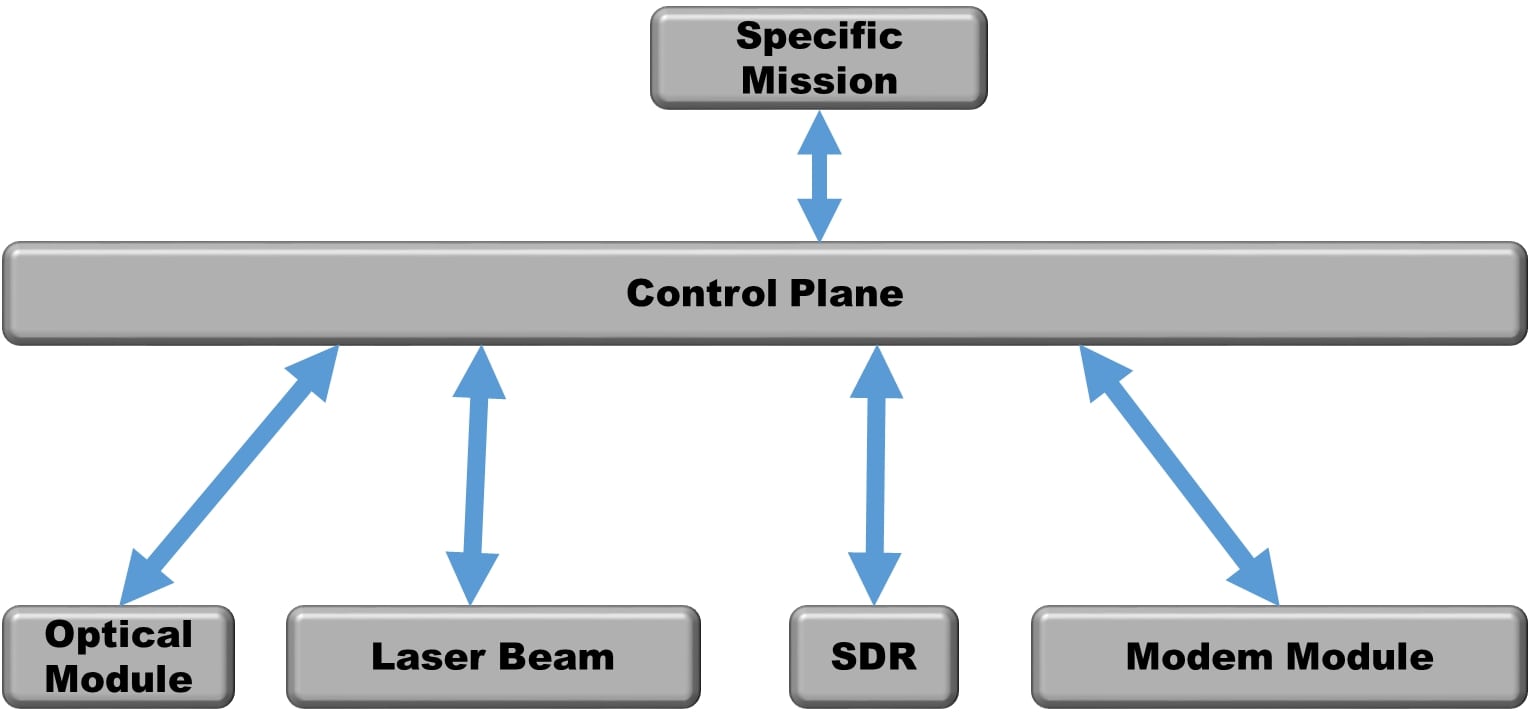}
		\subcaption{Mission-dependent SDN hierarchy align with SDR over specific element in IPN architecture.} 
	\end{subfigure}
	\caption{Integrating SDN on IPN} 
    \label{fig:sdn}
\end{figure}

The technologies we discussed in this paper still require lots of research and development before being enabled in a large scale project such as the IPN.
To provide 2-way laser communication and tackle bandwidth asymmetry, current technologies are in drastic need of miniaturization, not only for the sake of better functionalities but also for lowering the cost of deployment. 
One of the challenges to improve bidirectional link establishment time resides in improving the pointing methods. When a node points to another node, the receiving node should be able to infer the emplacement of the transmitting node from the light signal itself and respond by an uplink laser beacon to guide the transmitter, then redirect the laser beam to the exact location of the the receiving telescope.

In the IPN, deploying a new node takes several years, and systems work autonomously for decades. The possibilities for hardware failure should thus be minimized.
On Earth's computer networks, the current trend is to replace hardware elements with software. This new approach allows reducing deployment costs, while enabling us to update the internal mechanisms without physical access to the hardware. Two technologies are especially promising for the IPN. First, Software Defined Radio (SDR) delegates all the modulation operations to software modules, and use generic radio antennas for transmission. This architecture helps to achieve the desired miniaturization while limiting the number of points of failure on the hardware. As shown Figure~\ref{fig:sdn}, Software Defined Networks (SDN) separates the control plane from the data plane and allows to dynamically redefine the logical architecture of the network, allowing greater long-term flexibility.
In Figure~\ref{fig:sdn}, we demonstrate how to use SDN to control the equipments for a given mission. We apply SDN on the element level to reflect the requirements of the mission, for example control the transmit power, the beam width, and the RF band used depending of SDR.
\balance

The Interplanetary Transport Network (ITN) is a collection of gravitationally determined pathways through the Solar System that requires negligible energy for the spacecraft to follow. One instance of these pathways are the Sun-planet Lagrangian points.
These points give the opportunity to perform formation flying~\cite{leitner2004formation} or deploy constellations of spacecrafts to transmit data back and forth depending on their motion through the solar system with minimal maintenance.

Caching data is also an important issue to study. Placing previously requested information in temporary storage, or cache reduces demand on bandwidth and accelerates access to the most active data. In the case of human colonies on other planets, this functionality is vital for sharing the human knowledge pool between all planets. The technology to use for such functionality needs to be carefully discussed, as storage lifespan (which is already limited on Earth) may be severely reduced by the electromagnetic radiations present in deep space. Finally, a caching architecture designed for decades raises the question of the amount of storage to integrate for the system not to become obsolete. 
\section{Conclusion}

Space agencies and private organization have developed the technologies to conduct many missions for the purpose of exploring the universe, find alternative resources and broaden the science. Currently, the communication architectures supporting these missions are point-to-point and mission dependent.

This paper proposes an \textbf{evolutionary architecture} towards an Interplanetary Internet (IPN) to migrate from mission-centric architectures to a single common, scalable and reliable architecture.
Through this paper, we propose an evolutionary architecture based on the time-frame of future missions. The \textbf{Near-Term architecture} uses the available technologies and employ them to create an architecture that interconnect the current and near-future regions of interest (Mars and Moon). This architecture uses DTN and the suitable protocols to overcome the challenges enforced by deep space communication. The \textbf{Mid-Term architecture} addresses the growing demand to exchange huge amount of data between planets, especially after  colonizing Mars. This architecture's main evolution takes place in the spectrum architecture. Transmission transits from using RF communication to Laser communication, aligned with the foreseen evolution of hardware. Finally, for further reaching and continuous communication, we propose a \textbf{Long-Term architecture}. In this architecture, spacecraft are placed as relay nodes in space where they serve as repeaters to bridge the large distances, and overcome solar conjunction and signal attenuation.
We integrate DTN into each stage of our approach and propose a solution for multipoint communication and \textbf{autonomous routing} in space. We finally provided an integration scenario for sending a file from Jupiter to the Earth passing through Mars using every component of our architecture, and discussed implementation concerns regarding the feasibility of an IPN with current technologies.

With this paper, we hope to have provided a novel point of view for future IPN architectures, and set some foundations for an actual implementation within the next decades.
\newpage





\bibliographystyle{IEEEtran}
\bibliography{references}

\end{document}